# Social Echo Chambers in Quantum Field Theory: Exploring Faddeev-Popov Ghosts Phenomena, Loop Diagrams, and Cut-off Energy Theory

Yasuko Kawahata [†]

Faculty of Sociology, Department of Media Sociology, Rikkyo University, 3-34-1 Nishi-Ikebukuro,Toshima-ku, Tokyo, 171-8501, JAPAN.

ykawahata@rikkyo.ac.jp

**Abstract:** This paper presents an interdisciplinary approach to analyze the emergence and impact of filter bubbles in social phenomena, especially in both digital and offline environments, by applying the concepts of quantum field theory. Filter bubbles tend to occur in digital and offline environments, targeting digital natives with extremely low media literacy and information immunity. In addition, in the aftermath of stealth marketing, fake news, "inspirational marketing," and other forms of stealth marketing that never exist are rampant and can lead to major social disruption and exploitation. These are the causes of various social risks, including declining information literacy and knowledge levels and academic achievement. By exploring quantum mechanical principles such as remote interaction, proximity interaction, Feynman diagrams, and loop diagrams, we aim to gain a better understanding of information dissemination and opinion formation in social contexts. Our model incorporates key parameters such as agents' opinions, interaction probabilities, and flexibility in changing opinions, facilitating the observation of opinion distributions, cluster formation, and polarization under a variety of conditions. The purpose of this paper is to mathematically model the filter bubble phenomenon using the concepts of quantum field theory and to analyze its social consequences. This is a discussion paper and the proposed approach offers an innovative perspective for understanding social phenomena, but its interpretation and application require careful consideration.

**Keywords:** Social Media Dynamics, Quantum Theoretical Models, Online Social Bubbles, Cross-Disciplinary Approaches, Indefinite Metering Ghosts, Cut-off energy, Faddeev-Popov Ghosts, Anti-Ghost Field Influence, Loop Diagrams

## 1. Introduction

Filter bubbles tend to occur in digital environments and offline as well, targeting digital natives with very low media literacy and informational immunity. These are conditions that are extremely detrimental to the informational health of a society, and can lead to questions about the level of social trust, knowledge, and literacy in a given region or community, which in turn can trigger major social divisions, such as the risk of "social exploitation" in the medium to long term. In addition, stealth marketing and other forms of marketing have recently been gaining attention. In addition, in the aftermath of stealth marketing, etc., fake news and stealth marketing that are caused by things that do not exist and psychic sales, etc., are also rampant these days, and these can be an opportunity to create great social confusion and exploitation, and can also be the cause of the decline in information literacy, knowledge level, and academic ability in society, etc. They are also the cause of various social risks, such as a decline in the level of information literacy, knowledge, and academic achievement in society. In this paper, we consider these as social phenomena that occur regardless of online or offline, and furthermore, we hypothesize that these rumors and discourse are always flying around in a quantum manner, and our motivation is to solve them as a socio-physical method that applies quantum science. The approach of this paper is hypothetical and testable, but in the medium and long term, I would like to analyze the data with room for comparative verification, actual anomalies, and the situation of cases where filter bubbles have occurred, by comparing and verifying them.

One of the applications of relativity theory discussed in the context of quantum field theory is the discussion of remote interactions. There is also a discussion of proximity interactions. In this case, when considering the probabilities of divergence, convergence, and intermediate states, it is not only necessary to satisfy the laws of conservation of momentum and energy, but various states can be assumed as far as the uncertainty principle permits. For example, in terms of Feynman diagrams, phenomena such as opinion loops can be explained in terms of loop diagrams. It is a very interesting attempt to explain the discussion of remote interaction and proximity interaction in quantum field theory, especially the idea of a loop diagram using the Feynman diagram, by comparing it to phenomena occurring in society. In this case, we are taking the complex mathematical concepts of quantum field theory and replacing them with social phenomena that are more intuitive and easier to understand. When it is rigorously established apart from perturbation theory, etc., we call it the exact renormalization group equation. The renormalization group equation itself, which shows dependence on the truncation, does not include the energy of divergence because the modes in the neighborhood of the truncation are effective. However, if an effective action of a certain energy is given as an initial condition and integrated by the renormalization group equation, high-energy behavior can be calculated. In this case, the energy is analyzed when the coefficients of the initial condition are changed in a disconnected manner, so it is called a flow equation. In this case, the coefficients of the initial condition are fixed points, but the question is how much of the interaction terms between the quanta remain as convergent coefficients. In this case, what should be considered is that the converging coefficients form a hyperplane in the space of the overall coupling coefficients, which is a critical surface. Therefore, if the number of interaction operators is finite, the theory becomes renormalizable and asymptotically safe. It is also said that each quantum perturbation occurs and the degree of renormalization to electrons, charges, and masses in the case of divergence depends on the magnitude of the interaction. Discussions of perturbation theory in quantum field theory, in particular, the phenomenon of divergence and "renormalization" (renormalization) of charges and masses, using social phenomena as an analogy, is one way to make complex physics concepts more accessible. Proposing a mathematical model that applies quantum field theory concepts to analyze the phenomenon of filter bubbles in society is an ambitious attempt to cross the boundary between social science and physics. In the introduction of this paper, we aim to explore the mechanisms of filter bubbles in both digital and offline environments and analyze their social consequences. Specifically, we attempt to understand the propagation of information and the formation of opinions in social phenomena using concepts such as remote interaction, proximity interaction, Feynman diagrams, and loop diagrams in quantum field theory.

Social Background and Challenges of Filter Bubbles Filter bubbles in society are especially evident in individuals with low media literacy and information immunity. These filter bubbles can negatively impact informational health and lead to lower levels of social trust, knowledge, and literacy. In addition, they can cause social exploitation and fragmentation, and can cause major social disruption through phenomena such as stealth marketing and fake news.

## 1.1 Application of Quantum Field Theory Concepts to Social Phenomena

## 1.2 Introducing the Concepts of Remote Interaction and Proximity Interaction

The concepts of remote interaction and proximity interaction in quantum field theory are useful for understanding the propagation of information and opinion formation in society. Remote interaction is analogous to the phenomenon of distant people influencing each other through social media and the Internet. Proximity interaction, on the other hand, corresponds to face-to-face communication and interaction that occurs in physically close relationships. The concepts of Feynman diagrams and loop diagrams are well suited for understanding the circulation of opinions and self-referential feedback loops. This can be likened to the echo chamber phenomenon on social media. The attempt to apply concepts in quantum field theory to the analysis of social phenomena, especially filter bubbles, is an attempt to combine theories from physics with theories from the social sciences.

## 1.3 Remote Interaction

In quantum field theory, remote interaction refers to phenomena in which particles influence each other without physical contact. When this concept is applied to social phenomena, it can be likened to a situation in which physically distant people influence each other via the Internet or social media. It is suitable for describing the process by which information and opinions propagate across great distances to form social networks.

## 1.4 Proximity Interaction

Proximity interactions refer to interactions that occur when particles are physically close to each other. In its application to social phenomena, it represents interactions in direct human relationships and face-to-face communication. It helps explain the process of exchanging information and forming opinions between physically close people such as family, friends, and colleagues.

## 1.5 Feynman Diagram

A Feynman diagram is a graphic representation of the interaction of particles. It is an important tool for understanding complex quantum processes by providing a visual representation of the paths and interactions of particles. In its application to social phenomena, it can be used to diagram the flow of opinions and information and patterns of social interaction and to show how they interact with each other.

## 1.6 Loop Diagrams

A loop diagram is a type of Feynman diagram that shows how the interactions of particles form closed loops. This means that the particles return to their original state through a specific process. In social phenomena, it is suitable for modeling the "echo chamber" phenomenon, where information and opinions continue to circulate within a particular group or community.

### 1.7 Disconnection Energy

Disconnection energy is a concept introduced to ignore contributions above a particular energy scale. In quantum field theory, it is used to limit quantities that may diverge to infinity. In applications to social phenomena, it may be used to set limits or thresholds in the propagation of information or opinions. It thus serves as a parameter to control the frequency and intensity of interactions when assessing their impact on the formation or dissolution of filter bubbles.

### 1.8 Analysis of Filter Bubbles Using Mathematical Models

In this paper, we have attempted to examine some hypotheses and test them on a random number basis using a mathematical model based on the concepts of remote interaction and proximity interaction in quantum field theory. An agent-based approach is employed to simulate the dynamics of opinions through interactions between individuals in a social network. The model is based on parameters such as the opinions of the agents, the probability and frequency of interactions, and the flexibility and adaptability of opinion change. In the simulations, we observed the distribution of opinions, the formation of clusters, and the polarization of opinions under different conditions, and analyzed the mechanism of the formation of filter bubbles.

This is the case where fake news is a transition where the filter bubble becomes an eigenstate and finally only the bottom state affects the eigenstates that believe fake news as an eigenvalue. The logic of obtaining the above is such that the integrals and inferences of the paths along which the fake news occurred or the filter bubble was generated can be obtained.

Introducing the coordinate operator in the Heisenberg drawing, and further defining the corresponding operator in the Schrodinger drawing for the process of dissemination of information that should not exist, the state with the same eigenvalue is we further introduce a perturbation in which particles are stochastically amplified by time evolution, and here, by taking the limit of the infinite past using the time ordered product with Green's function, we obtain a transition formula in which only the ground state affects the eigenstate that believes the fake news as an eigenvalue. We decided to introduce a computational process by formulating the transition that affects the eigenstates that are In this paper, we also parameterized the transition to the state of believing fake news as eigenvalues as an influence on the ground state, and introduced the idea of scoring. In particular, the application of quantum mechanical concepts in the context of social science may provide new insights into understanding the impact of fake news and the formation of filter bubbles. Here we discuss the modeling of fake news and filter bubbles using Heisenberg and Schrödinger drawings.

### 1.9 Quantum Models of Fake News and Filter Bubbles

**Application of the Heisenberg Picture**

(1) **Introduction of the Coordinate Operator** $Q(t)$: In the Heisenberg picture, the time-dependent operator $Q(t)$ is crucial. This operator is used to represent social discourse and the flow of opinions and is utilized to model the impact of fake news.

(2) **Time Evolution**: The time evolution of the operator $Q(t)$ demonstrates how it is influenced by fake news. Over time, these effects change and impact social discourse.

**Application of the Schrödinger Picture**

(1) **Definition of the State Vector** $\Psi$: In the Schrödinger picture, the state vector $\Psi$ is essential. This vector represents the social state affected by the spread of fake news and fictional discussions.

**Influence on the Ground State**

In the ground state $|0\rangle$, the transition to a state that believes in fake news as an eigenvalue is calculated.

**Mathematical Model Calculation Process**

(1) **Introduction of Perturbation** $V(t)$: The perturbation $V(t)$ represents the impact of fake news.

(2) **Use of the Green's Function**: The Green's function $G(t, t')$ is used for the calculation of the time-ordered product.

(3) **Calculation of the Impact of Fake News**: The impact of fake news $\Psi(t)$ is calculated using the time evolution operator.

$$\text{Time evolution operator:} \quad U(t, t') = T \exp\left(-i \int_{t'}^{t} V(\tau) d\tau\right)$$

$$\text{Green's function:} \quad G(t, t') = -i \langle 0|T[Q(t)Q(t')]|0\rangle$$

$$\text{Impact of fake news:} \quad \Psi(t) = U^\dagger(t, -\infty) Q(t) U(t, -\infty)$$

**Modeling the Effect of Cutoff Energy**

The concept of cutoff energy is introduced to limit the frequency and intensity of interactions between agents. The aim of this paper is to mathematically model the phenomenon of filter bubbles in society using the concepts of quantum field theory and to analyze their social impact.

**Final Remarks**

In the aforementioned state, the eigenstate that believes in fake news is further affected by perturbations over time, leading to the emergence of vacuum bubble graphs.

$$\text{Feynman propagator:} \quad D_F(x, y) = \frac{1}{m^2 - ie}$$

# 2. Discussion of remote interaction and proximity interaction in quantum field theory (with a view to digital hygiene)

In the context of discussing hygiene issues in digital environments, there exists a discussion of remote interaction and proximity interaction. In this connection, we need to consider the convergence and divergence of information in the digital environment and probabilities in intermediate states. In addition, a variety of states are possible, taking into account the influence of the uncertainty principle. To understand these phenomena, analogies to the Feynman diagram will be discussed.

## 2.1 The concept of remote interaction

In the digital environment, it is common for users to exchange information and data from physically distant locations. In a remote interaction, energy and information are transferred to a remote point, and conditions that satisfy momentum and energy conservation laws are important. This remote interaction will be a relevant concept for communication technology and cloud computing,

## 2.2 The concept of proximity interaction

In a digital environment, users may share information in close proximity. In proximity interactions, devices and sensors are in close proximity and physical processes are involved. In this case, information is transferred through physical distance and contact, and momentum and energy conservation laws are also considered.

## 2.3 Convergence and Divergence Concepts

In a digital environment, information can be aggregated from different sources (convergence) or scattered in many directions (divergence). In convergence, information may be intensively combined and aggregated at a particular point. In divergence, information is diffuse and dispersed in many different directions. It will be important to strike a balance between convergence and divergence of information.

## 2.4 Intermediate States and the Concept of Probability

In the digital environment, intermediate states exist in the process of information transfer, and different states occur with varying probabilities. For example, the probability of an error during data transfer or the probability of information taking multiple paths would be considered.

## 2.5 The Uncertainty Principle

The effects of the uncertainty principle exist in the digital environment as well. It is difficult to simultaneously have perfect information about the location and velocity or path of transmission of information, and determination of one increases uncertainty about the other.

## 2.6 Loop Diagram Concept

The concept of a loop diagram can be used to describe the phenomenon of circular interaction of information in a digital environment, forming feedback loops. For example, when information is sent from a user to a device and then back to the user, the interaction is represented via a loop diagram.

Thus, by translating the concepts of quantum field theory into hygiene problems in the digital environment, we can better understand the various factors involved in information transfer and interaction. There is hope that this analogy will be useful in exploring the complexities of information flow and interaction in the digital environment.

Each of the above is said to generate a quantum perturbation, and the degree of renormalization into electrons, charge, and mass when divergence occurs depends on the magnitude of the interaction.

In hygiene problems in the digital environment, phenomena corresponding to quantum perturbations occur in relation to the handling of digital information and data transmission. Depending on the magnitude of information handling and interaction, the behavior and mass (importance and impact of information) of digital entities (e.g., devices, applications, users, etc.) associated with electrons and charges can be said to differ. The following could be explained using this analogy. In a digital environment, different digital entities interact with each other when it comes to the interaction perspective of digital information. This corresponds to interaction in quantum mechanics, where energy and resource inputs occur in the transfer and processing of information. 2. As for the magnitude of the interaction, the magnitude of the information interaction depends on the number of digital entities involved in the interaction and the complexity of the information. Some information may exist independently with little interaction with other information. On the other hand, a high degree of interaction may occur, and information may affect many processes and entities. With respect to divergence and convergence, in the digital environment, there are cases where information interactions lead to information aggregation (convergence) and cases where information spreads in many directions (divergence). For example, when a particular topic is widely shared on social media, information is diverging. On the other hand, encryption of information, such as security protocols, facilitates information convergence. Thinking in terms of digital entities as alternatives to electrons and charges, information or digital entities are the agents of interaction in a digital environment, whereas electrons and charges are physical particles. Digital entities vary in their handling and influence of information; some digital entities may be involved with a lot of information while others may be involved with a limited amount of information. Furthermore, mass, when considered by analogy with mass, represents the importance or quality of a physical object, but in a digital environment, the quality and importance of information and data fluctuates. Important information can be considered to have a high mass, and that information can have a significant impact in interactions. In the digital environment, there is a need for discussion and coordination of information interactions. Discussions regarding the divergence and convergence of information, the quality and importance of information, and the development of appropriate rules and policies to ensure the sanitation of the digital environment. This analogy can be used to consider the interaction and quality of information, the convergence and divergence of information, and information hygiene in the digital environment. We believe that information interaction is important in the digital environment and that understanding and coordinating this interaction will contribute to ensuring a healthy information environment in the digital society.

## 3. Discussion:Renormalizable Theory

Furthermore, if the coupling constant has no dimension in the natural unit system with h=c=1, it can be called a "renormalizable theory". Thus, scattering amplitudes and vacuum structures can also be derived when considering the effective action in the field. In this case, the renormalization group at divergence gives the effect of the cutting energy and the process of redefining it to observable values. In this case, the concept of "renormalizable theory" is replaced by the problem of hygiene in the digital environment. In the digital environment, the concept of renormalization is also useful in relation to information handling and sanitary conditions. Introducing the concept of renormalizable digital environments, a renormalizable digital environment is one in which the coupling and interaction of information and digital entities has no dimensions and information can be processed at different scales. In such digital environments, information and processes are scaled and there are universal properties that are independent of specific scales. Introducing the concept of scattering amplitude and vacuum structure, information transfer and interaction also occur in digital environments. In this case, scattering amplitude indicates the degree and effect of information interaction, while vacuum structure represents blank or unused information within the digital environment. If information transfer and interaction occur at different scales, the scattering amplitude and vacuum structure will vary accordingly.

Introducing the concept of renormalization groups and cut energy, renormalization groups are processes that account for the scale dependence of information processing in the digital environment. Cutting energy represents the energy cutoff at the scale of information, which affects the selection of information to be discarded. The renormalization group considers the effects of information scaling and energy and redefines them into observable values and sanitary conditions.

Introducing the idea of the health of the digital environment, the health of the renormalizable digital environment would be related to managing the scale dependencies and energy effects of information and maintaining sanitary conditions. Through the renormalization group, information processing and interactions within the digital environment will be coordinated, limiting inappropriate information and excessive interactions.

Once the concept of improving the digital environment is introduced, steps can be taken to improve the health and hygiene of the digital environment using a renormalizable approach. Information scaling and energy cutoffs will be adjusted to facilitate appropriate information transfer and interaction. This makes the digital environment more effective and sustainable.

This analogy underscores the importance of information processing and hygiene in the digital environment and suggests that the concepts of the renormalizable theory can be applied to improve the health of information handling and interactions. In the digital environment, scaling of information processing and management of energy effects could contribute to the creation of a sustainable digital society.

## 4. Discussion:Renormalization group

Scattering amplitudes and vacuum structures can also be derived when considering effective actions in quantum fields. In this case, the renormalization group at divergence gives rise to a process that redefines the observable with the effect of the truncation energy. The effect of the cutting energy is called the effective mean action, which generates a very large mode effect. When the equation is rigorously established without any perturbation theory, it is called the exact renormalization group equation.

Translating this argument to the hygiene problem in the digital environment, if we think in terms of effective action in the digital environment, information processing and interactions between digital entities are relevant to the handling of data and the execution of processes in the digital environment. This interaction controls events and behavior within the digital environment.

When considered in terms of scattering amplitude and vacuum structure, information is scattered in the digital environment as well as in the process of data change and communication due to information transfer and interaction. Vacuum structure represents the state of unused resources and data within the digital environment and is associated with efficient use of information. When considered in terms of renormalization groups and cutting energy, the concept of renormalization groups applies when information processing and interactions occur at different scales in the digital environment. Cutoff energy represents an energy cutoff in information processing, limiting the processing of unnecessary information or excessive data. The process of renormalization groups contributes to the redefinition and optimization of information within the digital environment. When considered in terms of effective mean action and mode effects, the effect of cutoff energy manifests itself as an effective mean action within the digital environment. This means that the digital environment functions properly without unnecessary data or excessive information processing. Some large mode effects are important to avoid data concentration and overload. Considered in terms of exact renormalization group equations, the renormalization group equations are applied in rigorous analysis of sanitation problems in the digital environment. This equation is used to mathematically model the effects of data scaling and processing to maintain the health and sanitation of the digital environment.

This analogy highlights information processing and data hygiene issues in the digital environment, and suggests that the concepts of the renormalizable theory can be applied to improve data management and information security, and that the hypotheses and testability of the theory can be tested. Understanding the renormalizable group equation on this definition will be important to ensure that the digital environment maintains proper data processing and sanitary conditions.

## 5. Discussion:Disconnect dependence, Flow equations

When the equation is rigorously established apart from kinetic theory, etc., it is called the exact renormalization group equation. The renormalization group equation itself, which shows the dependence on the truncation, is valid for modes near the truncation and does not include divergence energy. However, if an effective action of a certain energy is given as an initial condition and integrated by the renormalization group equation, high-energy behavior can be calculated. In this case, the energy is analyzed when the coefficients of the initial condition are changed in a disconnected manner, so it is called a flow equation. In this case, the coefficients of the initial condition are fixed points, but the question is how much of the interaction terms between the quanta remain as convergent coefficients. In this case, what should be considered is that the converging coefficients form a hyperplane in the space of the overall coupling coefficients, which is a critical surface. Therefore, if the number of interaction operators is finite, the theory becomes renormalizable and asymptotically safe.

Let us also replace this argument with the hygiene problem in the digital environment. With respect to strict hygiene conditions in the digital environment, the processing of data and the interaction of information in the digital environment must also meet certain hygiene conditions. These hygiene conditions would be replaced by the idea of ensuring the accuracy, security, and efficiency of information within the digital environment.

With regard to the discussion of disconnect dependence and the digital environment, "disconnect dependence" refers to the impact of certain conditions or criteria on the effectiveness of data processing and information management within the digital environment. Disconnections in the digital environment will play an important role in the collection, storage, and transmission of data.

In the digital environment with respect to flow equations and initial conditions, one may start with certain initial conditions and analyze the data processing and information transmission process in detail. In this case, flow equations are used to predict the behavior of the data as the initial conditions change. As the coefficients of the initial conditions change, the specific behavior and performance of the data can be analyzed.

It is then important to determine the convergence coefficients and whether a particular data processing process or information interaction satisfies the convergence conditions in the digital environment on the critical surface. Convergence coefficients relate to data quality and processing stability within the digital environment and form critical surfaces. If the critical surface is exceeded, there will be potential problems with data quality and information security. With respect to asymptotically secure digital environments, ultimately, when data processing and information interaction within the digital environment converge and reach a secure and efficient state, this can be called an asymptotically secure digital environment. Effective information processing is achieved because information is handled appropriately and data quality is maintained within the digital environment.

This analogy will emphasize the soundness and security of data in the digital environment and suggest that the concepts of the Theory of Retrievability can be applied to improve data management and information security.

# 6. Conclusion: Modeling Filter Bubbles with Quantum Fields in Agent-Based Modeling

**Modeling Filter Bubbles with Quantum Fields in Agent-Based Modeling**

The attempt to model the phenomenon of filter bubbles in society and consider the effects of cutoff energy is an intriguing fusion of theories from social science and physics. Here, we consider agent-based modeling as one approach to model filter bubbles.

### Overview of the Model

Agents (individuals) each have their own opinions and beliefs and interact through platforms such as social media.

Agents have an affinity for specific opinions or beliefs, leading to a higher frequency of interaction with other agents who share similar opinions (formation of filter bubbles).

### Parameters

Initial distribution of opinions or beliefs among agents.

Probability and frequency of interactions between agents.

Flexibility and adaptability of opinion changes.

### Computational Process

(1) At the start of the simulation, opinions are assigned to each agent randomly or according to a specific distribution.

(2) At each time step, agents interact stochastically with other agents, determining whether their opinions change or become reinforced.

(3) Over time, opinion clusters form, and filter bubbles are observed.

### Modeling the Effects of Cutoff Energy

Cutoff energy serves as an upper limit to prevent the divergence of theories, akin to the concept in physics. In social modeling, this concept can be incorporated as "limits on opinion changes" or "constraints on interactions."

### Agent State (Opinion) Model

Each agent $i$ holds an opinion $o_i$ represented as a real number (e.g., ranging from -1 to 1).

Initially, agent opinions are assigned randomly or according to a specific distribution.

### Interaction Model

When agents $i$ and $j$ interact, their opinions are updated as follows:

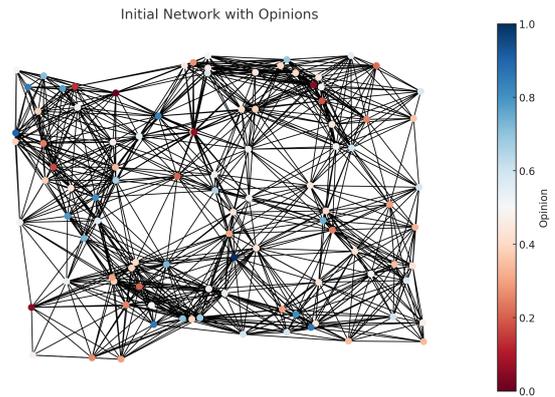

Fig. 1: Initial Network with Opinion

$$o_i^{(new)} = o_i + \alpha(o_j - o_i), \quad o_j^{(new)} = o_j + \alpha(o_i - o_j)$$

Here, $\alpha$ is a parameter representing the rate of opinion adaptation and takes values between 0 and 1.

### Parameters

Adaptation rate $\alpha$: A higher value means that agents are more strongly influenced by others' opinions.

Interaction probability $p$: The probability of any two agents interacting.

Cutoff energy $E_{cut}$: An upper limit set on the number of interactions or the extent of opinion changes.

### Computational Process

(1) Simulation Loop:

At each time step, randomly selected agent pairs interact with a probability $p$.

Agents' opinions are updated following the above rules.

If the number of interactions between agents reaches $E_{cut}$, further opinion changes are restricted.

Results depict what appears to be two states of a network with nodes representing agents and edges representing interactions between them. Each node has an associated opinion value, indicated by its color on a scale from red (negative opinion) to blue (positive opinion).

## 6.1 Initial Network with Opinions (First Image)

In the initial state, the network's opinions are more evenly distributed across the spectrum. We can observe:

1. Diversity of Opinions: There's a mix of opinions, as indicated by the range of colors. This suggests a starting point where agents hold a variety of stances, and no single opinion dominates.

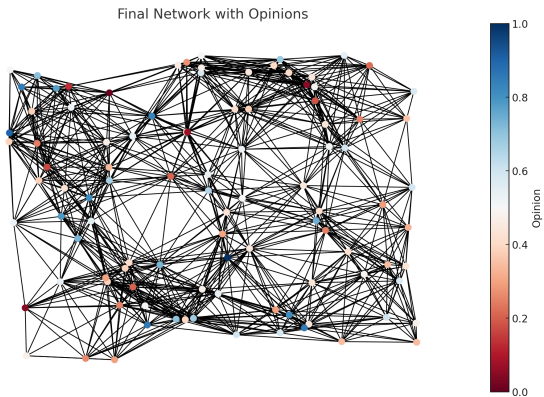

Fig. 2: sFinal Network with Opinions

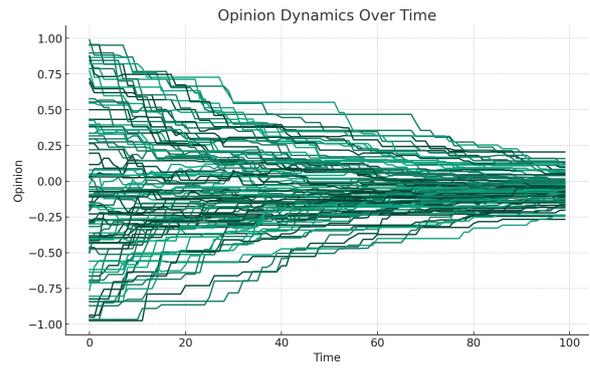

Fig. 3: Opinion Dynamics Over Time

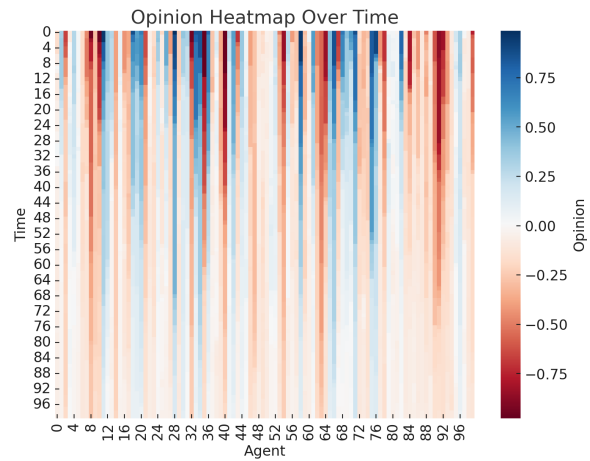

Fig. 4: Opinion Heatmap Over Time

2. Weak Clustering: There are no strong clusters visible, which might imply that the network is in an early stage of opinion formation, or that the adaptation rate ($\alpha$) and interaction probability ($p$) have been set in a way that promotes diversity.

3. Potential for Polarization: Given that the opinions are spread out, there is a potential for the network to evolve into more polarized clusters, depending on how the agents interact and adapt to each other's opinions.

### 6.2 Final Network with Opinions (Second Image)

The final state shows a change in the opinion distribution:

1. Opinion Shifts: Some agents have shifted their opinions, indicated by the change in color. There seems to be a slight movement towards the blue end of the spectrum, which might indicate an overall shift in opinion towards the positive side.

2. Formation of Clusters: There appears to be more clustering, with groups of agents sharing similar opinions. This suggests that over time, agents have influenced each other, leading to a more pronounced clustering effect.

3. Extremization of Opinions: While not all agents have moved to the extremes, there is a visible concentration of like-minded agents. This can be a sign of extremization, where agents within a cluster reinforce each other's opinions.

### 6.3 Considerations

Adaptation Rate ($\alpha$): If $\alpha$ is high, agents quickly adopt the opinions of those they interact with, leading to rapid cluster formation. In the final state, the presence of more defined clusters may indicate a higher $\alpha$. Interaction Probability ($p$): A higher $p$ increases the likelihood of agents interacting and influencing each other. If $p$ is high, we expect the network to reach a consensus or form distinct clusters more quickly.

Comparing the initial and final states, it appears that the agents' opinions have undergone significant shifts and are now more clustered, indicating a process of social influence and potential echo chamber formation. The dynamics of this network can be influenced by various factors, including the underlying structure of the network, the adaptation rate, and the probability of interaction. The evolution of opinions and the formation of clusters in such a network could serve as a microcosm for understanding how opinions spread and solidify in society, particularly in the context of the formation and effects of filter bubbles.

Results regarding the distribution of opinions, formation of clusters, and polarization of opinions, taking into account the adaptation rate ($\alpha$) and the probability of interaction ($p$) between agents.

### 6.4 Initial Network with Opinions

The first graph represents the initial state of a social network where each node (or agent) has an opinion value indicated by the color. The opinions range from 0 to 1, where the color scale from red to blue indicates the spectrum of opinions. The network is densely connected, meaning each agent is influenced by many others.

From the graph, we can observe a diverse distribution of opinions without any clear clusters. This suggests that, initially, there is no significant polarization or echo chamber effect. The variation in opinion colors indicates a mixed society without predominant opinion groups.

### 6.5 Final Network with Opinions

The second graph shows the network after interactions based on the rules defined by the adaptation rate and interaction probability. Here, we can see that the opinions have shifted, and there seems to be a more uniform opinion across the network, indicated by the prevalence of a particular color. This suggests that agents have influenced each other towards a consensus or a dominant opinion. This could be due to a high adaptation rate where agents strongly adjust their opinions towards those they interact with.

Clusters seem to form where nodes with similar opinions are now more closely aligned. This can indicate the formation of echo chambers or opinion clusters where agents mostly interact with others who have similar opinions, reinforcing their existing views.

### 6.6 Opinion Dynamics Over Time

The third graph shows the opinion dynamics over time. Each line represents the opinion trajectory of an individual agent. We can see that over time, the opinions converge, indicating a collective shift towards certain opinion values. This is characteristic of systems with a high adaptation rate, where individual opinions become more aligned with the group consensus as time progresses. The absence of divergent lines towards the end suggests a strong convergence of opinions and possibly a high interaction probability.

### 6.7 Opinion Heatmap Over Time

The fourth graph is a heatmap showing the evolution of each agent's opinion over time. The color intensity indicates the strength of the opinion. It is clear from the heatmap that there is a trend towards opinion convergence over time. This convergence could result from a combination of a high adaptation rate and a high probability of interaction, which together facilitate the rapid spread and adoption of certain opinions within the network.

In summary, the provided graphs suggest that over time, the agents in the network have influenced each other significantly, leading to opinion convergence. This process is possibly driven by a high adaptation rate where agents are highly receptive to the opinions of their peers, and a high probability of interaction, which ensures frequent exchanges between agents. The result is a network that moves from a state of diverse opinions to one that is more homogeneous,

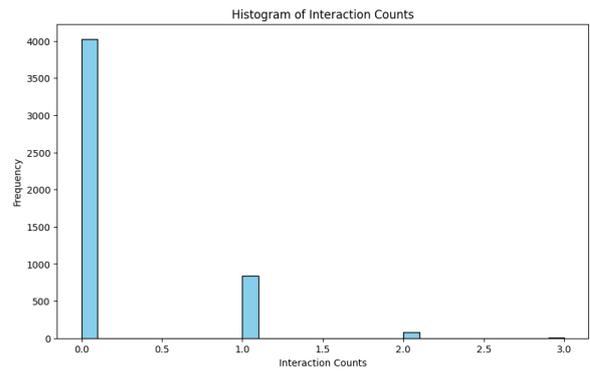

Fig. 5: Histogram of Interaction Counts

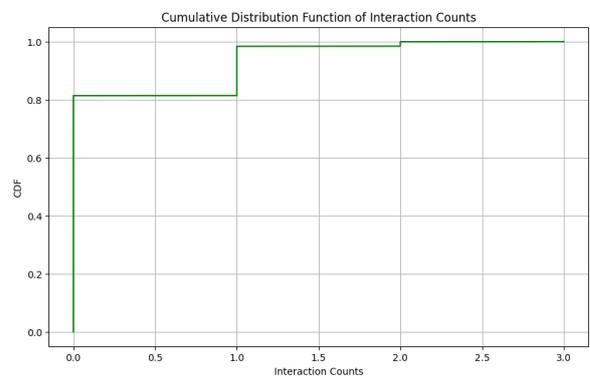

Fig. 6: Cumulative Distribution Function of Interaction Counts

with the potential for the formation of opinion clusters or echo chambers.

(1) **Opinion Distribution:**

If the network graphs show a variety of colors representing opinions, a spread of opinions at the beginning, converging towards consensus or diverging into distinct clusters over time, would be indicative of how opinions evolve. A more uniform color distribution in the final network would suggest consensus, whereas distinct color clusters would indicate polarization.

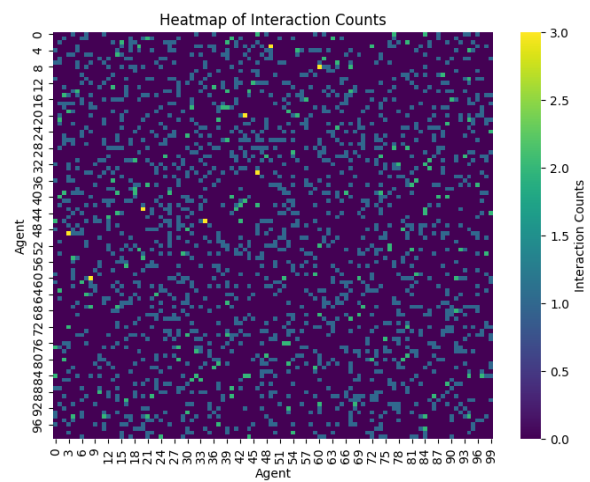

Fig. 7: Heatmap of Interaction Counts

(2) **Cluster Formation:**

In a social network graph, clusters would appear as groups of nodes more closely connected with each other than with the rest of the network. The initial network might show a random pattern, whereas the final network may have clear clusters if opinions have polarized. In an opinion dynamics model, clusters would indicate groups of agents that have aligned opinions, possibly due to strong local interactions or echo chamber effects.

(3) **Opinion Polarization:**

Over time, if the graphs show that the agents' opinions are becoming more extreme (i.e., moving towards the extremes of the opinion scale rather than staying near the center), this would suggest polarization. This could be evidenced by a bimodal distribution of opinions in the final network or by most agents having opinions near the extremes of the color scale.

### 6.8 Considering the adaptation rate $\alpha$

A higher adaptation rate implies that individuals change their opinions more readily based on their interactions. This could lead to faster consensus or more pronounced clusters, depending on whether the interactions tend to be reinforcing or diversifying.

### 6.9 Considering the probability of interaction $p$

If $p$ is high, agents interact frequently, potentially leading to rapid consensus or strong polarization. A lower $p$ might slow down opinion changes and lead to a more mixed final opinion state.

### 6.10 For the Histogram of Interaction Counts

This graph likely shows how often agents interact. A high frequency at lower interaction counts would suggest many agents don't interact often, which could slow opinion convergence or polarization.

### 6.11 For the Cumulative Distribution Function (CDF) of Interaction Counts

The CDF provides the probability that an agent's interaction count is less than or equal to a certain value. A steep curve suggests that most agents have a low number of interactions, and few have more, which again points to limited interaction across the network.

### 6.12 For the Heatmap of Interaction Counts

This would visually represent the interaction frequency between each pair of agents. In a model where interaction frequency influences opinion dynamics, clusters in the heatmap could correspond to opinion clusters in the network, indicating that frequent interactions lead to opinion alignment.

In summary, analyzing these graphs could reveal how strongly agents influence each other and how opinions spread and evolve in the network. If the network experiences strong polarization, it suggests that the underlying dynamics—be they high adaptation rates, frequent interactions, or specific patterns of connectivity—are conducive to forming opinion clusters rather than a consensus. Conversely, if the final network shows a more homogeneous opinion distribution, it might suggest that the dynamics encourage convergence to a common opinion or that the interaction strength (possibly modulated by the cutoff energy) is not sufficient to cause polarization.

## 7. Conclusion: Exploration of Flow Equations and Cutoff Functions

Let's delve further into the discussion about flow equations, especially when considering perturbation theory, cutoff patterns, and the issue of indefinite metrics associated with ghosts. Various types of cutoff energy patterns should be considered in the context of cutoff functions. In the case of perturbation theory with indefinite metrics due to ghosts, phase transitions do not occur. In such cases, when proceeding with quantum analysis after York decomposition of second-order symmetric tensors for each spin component, it becomes necessary to determine the propagator function of gravitons for BRST transformations. Additionally, the presence of the Faddeev-Popov ghost term (referred to as FP ghost) is essential. It's worth noting that convergence tends to be easier when negative ghosts are present (applicable to spin-1 cases only).

### 7.1 Cutoff Functions and Renormalization Group Flow

Considering cutoff functions leads to Legendre transformations, where the behavior of the renormalization group flow can exhibit two scenarios: one where it diverges towards infinity, resulting in a Landau pole, and another where it remains finite, leading to fixed points. In the latter case, the system becomes renormalizable. Moreover, this approach can help explain the concepts of limit cycles and ergodicity in opinion dynamics. In some scenarios, loop expansions or local potential approximations may be detected in opinion trajectories.

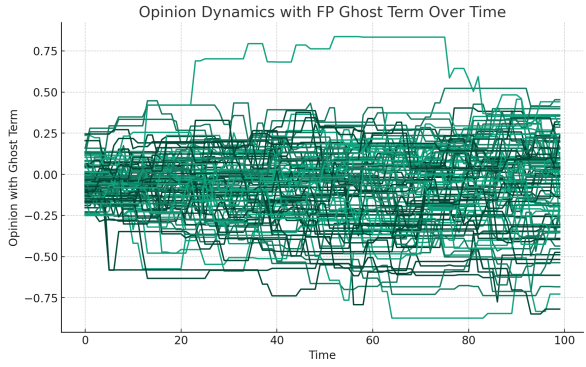

Fig. 8: Opinion Dynamics Final Network with Opinions and FP Ghost Term

## 7.2 Types of Cutoff Functions

Cutoff functions can be categorized into various types:

### 7.2.1 Type Ia Cutoff (Maximally Symmetric Space)

In Type Ia cutoff, anomalies known as "anomalous dimensions" can be observed. FP ghosts are also detectable and traceable.

### 7.2.2 Type Ib Cutoff (Using York Decomposition)

Type Ib cutoff involves the use of York decomposition, allowing the verification of an ultraviolet (UV) fixed point in four dimensions, with $G > 0$.

### 7.2.3 Type II Cutoff

Type II cutoff exists, but a pure cutoff is primarily applicable in Type III cutoff. In Type III cutoff, the flow equations introduce factors $\delta_t R_k$ and $R_k$, allowing the extraction of energy from modes in the cutoff neighborhood.

This discussion provides insights into the application of cutoff functions and renormalization group flow to understand opinion dynamics. The various types of cutoff functions offer different perspectives on the behavior of the system, shedding light on its renormalizability and the presence of anomalous dimensions and FP ghosts.

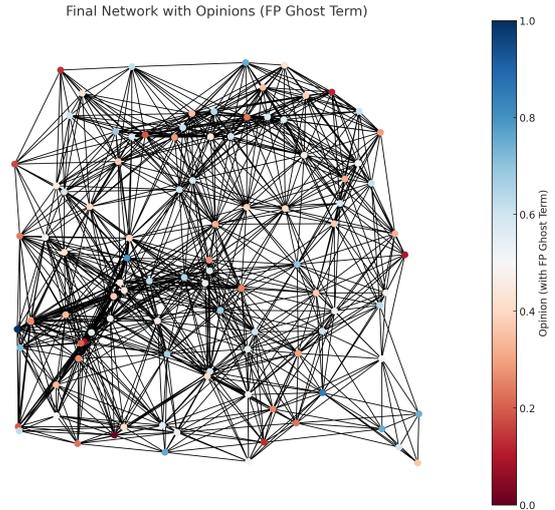

Fig. 9: Final Network with Opinions and FP Ghost Term

## 7.3 Opinion Dynamics Over Time with FP Ghost Term

Results which appear to include graphs related to opinion dynamics in a social network, with a particular focus on the impact of misinformation or biased information as represented by a "FP Ghost Term". we will consider the opinion distribution, cluster formation, opinion polarization, and the influence of non-physical elements like misinformation or bias.

**Opinion Divergence**: The introduction of the FP Ghost Term seems to increase the spread and divergence of opinions over time, indicating that misinformation or biases can lead to a wider range of beliefs within a network.

**Clustering**: If the network initially had clustered opinions, the FP Ghost Term might have disrupted these, leading to a more chaotic distribution of opinions as agents are influenced by non-physical elements.

**Polarization**: The presence of non-physical factors can exacerbate polarization if they reinforce existing biases or introduce new divisive elements into the discourse.

## 7.4 Final Network with Opinions

**Opinion Distribution**: The final state of the network could show a complex distribution of opinions where the FP Ghost Term may have led to the creation of subgroups with distinct beliefs or increased the isolation of certain agents.

**Cluster Formation**: The FP Ghost Term may prevent the formation of consensus clusters or lead to the formation of more, smaller clusters, indicating fragmented communities with entrenched viewpoints.

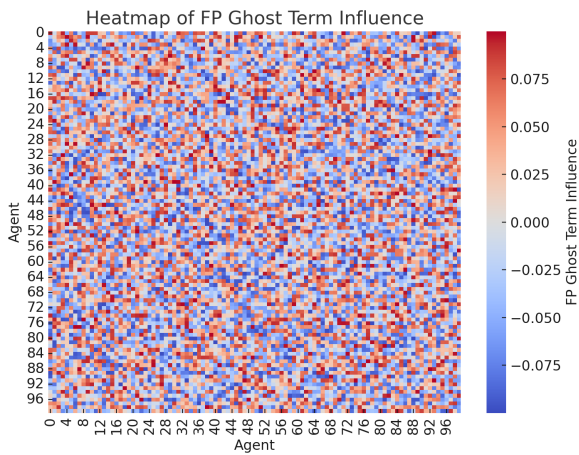

Fig. 10: Heatmap of FP Ghost Term Influence

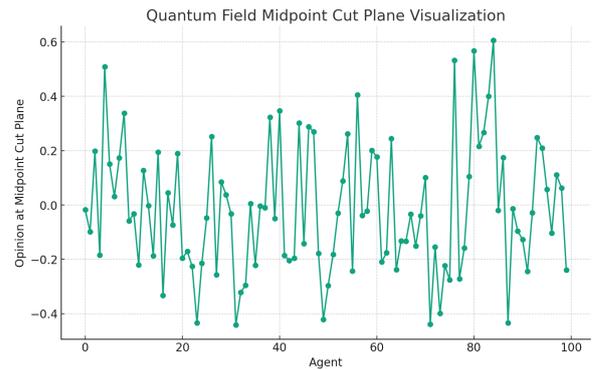

Fig. 11: Opinion at Midpoint Cut Plane

**Polarization**: If the network ends with agents having extreme opinions, it suggests that misinformation or bias can lead to a more polarized society, where middle-ground opinions become less common.

### 7.5 Interaction Counts and the FP Ghost Term

Considering the interaction counts, the FP Ghost Term can be thought of as an agent that interacts randomly with others, injecting misinformation or biased perspectives. It can serve as a source of noise in the system, leading to unexpected shifts in opinion dynamics. The specific impact on interaction counts would depend on the underlying rules of the network dynamics and how the FP Ghost Term is implemented (e.g., frequency of interaction, strength of influence, etc.).

The analysis of such a system would involve looking at how the introduction of the FP Ghost Term changes the stability of the network's opinion states, the time it takes for the network to reach equilibrium (if it does), and the nature of that equilibrium. This can be compared to a control scenario without the FP Ghost Term to isolate its effects.

It's important to note that without direct access to the specific graphs and data, the analysis is based on typical outcomes observed in opinion dynamics models when external influences are introduced. The actual graphs and numerical data would provide essential insights into the precise nature and extent of these effects.

### 7.6 Opinion Dynamics with FP Ghost Term Over Time

This graph shows the fluctuation of opinions over time with the introduction of the FP Ghost Term, which might represent the influence of non-physical elements like misinformation or bias. The presence of the FP Ghost Term seems to introduce additional noise or volatility in the opinion dynamics, which could represent the unpredictable nature of how misinformation or biases can affect public opinion over time.

### 7.7 Final Network with Opinions (FP Ghost Term)

This network visualization with the FP Ghost Term presents the final distribution of opinions. The color gradient indicates the range of opinions with the influence of the FP Ghost Term. If the term introduces non-physical effects like misinformation, the resulting polarization or consensus could be based on such distorted information flows.

These graphs suggest that the introduction of non-physical factors can have a significant and possibly destabilizing effect on opinion dynamics, leading to a more polarized or fragmented society, especially if these factors are allowed to propagate through the network unchecked.

### 7.8 Analysis of Graphs (Cut-off Energy and Bias)

This plot could represent the snapshot of opinions at a certain point in the quantum field model. The significant variance across agents' opinions suggests a diverse set of views or a state of disagreement within the network.

### 7.9 Heatmap of FP Ghost Term Influence

The heatmap shows the variation of the FP Ghost Term's influence across different agents over time. Red and blue patches indicate positive and negative influences, respectively. The pattern suggests that misinformation or bias (represented by the FP Ghost Term) is widespread and varies greatly between different agents.

From these observations, it can be inferred that misinformation or bias is not uniformly spread across the network and can lead to clusters of highly influenced or less influenced individuals. The cut-off energy could be a measure to prevent the spread of such non-physical factors, but its effectiveness

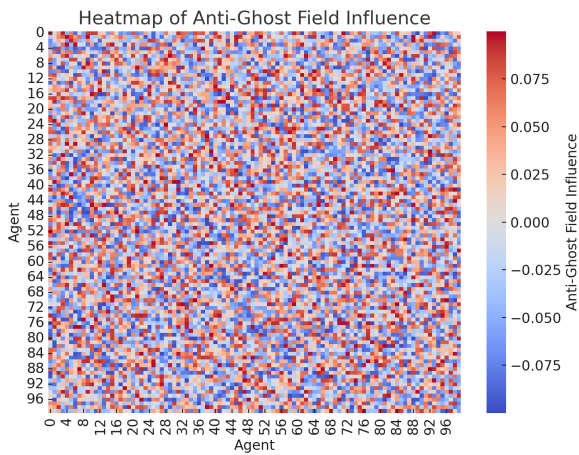

Fig. 12: Heatmap of Function of Anti-Ghost Field Influence

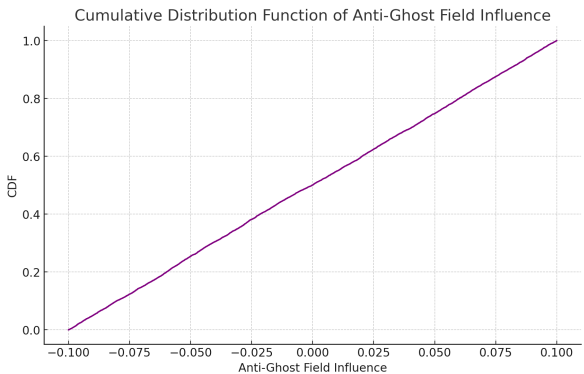

Fig. 13: Cumulative Distribution Function of Anti-Ghost Field Influence

would depend on the network's structure and the strength of the cut-off.

To conclude, the provided graphs depict a complex system where opinions are influenced by various factors, including interaction probabilities, misinformation, biases, and mechanisms like cut-off energy. These factors together shape the opinion dynamics, leading to the formation of clusters, the spread or containment of misinformation, and the polarization of views within the network.

### 7.10 Anti-Ghost Field Influence Analysis

(1) **Heatmap of Anti-Ghost Field Influence**

> **Distribution of Influence**: This heatmap likely illustrates the variation of the Anti-Ghost Field's influence across different agents in the network. The color-coded values may indicate the strength and direction (positive or negative) of the influence.
>
> **Patterns**: Look for clusters of similar colors indicating areas where the Anti-Ghost Field has a consistent impact. An even spread of colors might suggest that the influence is randomly distributed without clear patterns.
>
> **Intensity**: The depth of color might show the intensity of the Anti-Ghost Field's influence. Deep reds or blues could indicate strong influence, while lighter shades may suggest a weaker effect.

(2) **Cumulative Distribution Function of Anti-Ghost Field Influence**

> **Overall Influence**: This graph is probably a CDF that demonstrates the cumulative probability of the Anti-Ghost Field's influence being less than or equal to various values. It would be used to understand the proportion of agents affected by different intensities of influence.
>
> **Symmetry**: If the CDF is symmetrical around zero, it might indicate that the Anti-Ghost Field's influence is balanced between positive and negative impacts across the network.
>
> **Skewness**: Any skew to the left or right could imply a predominance of negative or positive influence, respectively.

For an in-depth interpretation, it would be necessary to consider the following factors:

> **Context of the Anti-Ghost Field**: Understanding what the Anti-Ghost Field represents in the model (e.g., misinformation, bias, external influences) is crucial. This context will inform how one interprets the distribution and impact of this field.
>
> **Network Dynamics**: The influence of the Anti-Ghost Field should be considered in the context of the network's dynamics. How do agents interact, and how might these interactions be affected by the field?
>
> **Potential for Opinion Change**: Analyze the potential for opinion change due to the Anti-Ghost Field. Are there agents or regions within the network that are more susceptible to change?
>
> **Mitigation Strategies**: If the Anti-Ghost Field represents undesirable influences, the heatmap and CDF can help identify where to target interventions or countermeasures.

## 8. Conclusion: Faddeev-Popov (FP) Ghost Term

In the context of digital quantum fields, we introduce a simulation-based discussion on anomalies related to information integrity, such as hacking and fake news. It is crucial to consider both remote interactions and close interactions in the digital environment. Users can share information while being physically close to each other in close interactions, where devices and sensors are nearby, and physical processes are involved. In this case, factors like physical distance, contact, momentum, and energy conservation come into play. In digital environments, users commonly exchange information or data from physically distant locations through remote interactions. In remote interactions, energy and information are transmitted to distant points, where conditions like momentum and energy conservation are essential. However, the introduction of the Faddeev-Popov (FP) ghost term can be problematic and indefinite in nature, raising concerns in digital hygiene, including the spread of fake news and filter bubbles.

## 8.1 Introduction of FP Ghost Term

The FP ghost term is a virtual field that arises in gauge theories during the process of gauge fixing. In social models, "ghosts" are used to represent hidden elements that affect the propagation of opinions, such as misinformation or biases.

## 8.2 Mathematical Representation

The mathematical representation of the FP ghost term in gauge theories is typically given as follows:

$$S_{FP} = \int d^4x\, \bar{c}^a \mathcal{M}^{ab} c^b$$

Here, $\bar{c}^a$ and $c^b$ are ghost and anti-ghost fields, respectively. $\mathcal{M}^{ab}$ represents the operator (e.g., FP matrix) that arises during gauge fixing.

## 8.3 Application to Social Models

### 8.3.1 Analogy of Ghost Term

In social models, the ghost term is introduced to represent the influence of non-physical elements on the propagation of opinions.

### 8.3.2 Application in Equations

In the context of social models, we apply the ghost term as follows:

Agent $i$'s opinion $o_i$ corresponds to the ghost field. The FP matrix represents social factors (e.g., biases or misinformation) affecting the propagation of opinions among agents.

### 8.3.3 Calculation Process

We include the FP ghost term in the interactions between agents and consider its influence on the propagation of opinions. This influence is added to the opinion update rules and affects how agents' opinions change.

### 8.3.4 Implementation Example

During simulations, we consider the influence of the FP ghost term when updating the opinions of agents. For example, if agents $i$ and $j$ update their opinions as follows:

$$o_i^{(new)} = o_i + \alpha(o_j - o_i) + \gamma G_{ij}$$

Here, $G_{ij}$ represents the influence of the FP ghost term, and $\gamma$ is the strength of that influence.

This model helps us understand the impact of hidden elements, such as misinformation or biases, in social interactions, but further validation may be necessary.

# 9. Conclusion:Quantum Field Discussion through Legendre Transformation and Renormalization Group

In this section, we introduce the discussion of anomalies in digital quantum fields, specifically addressing issues like hacking, fake news, and other forms of information manipulation in a simulated context. An essential aspect of this discussion revolves around the digital environment, where users may share information in both remote and proximal interactions. In proximal interactions, devices and sensors in close physical proximity play a role in the information exchange.

We will elucidate the specific computational processes involved in introducing the FP ghost term in social models, followed by Legendre transformation and analysis of the renormalization group. This process aims to apply the complex concepts of quantum field theory in the context of social sciences.

## 9.1 Introduction of FP Ghost Term

1. Modeling the FP Ghost Term: Each agent's opinion $o_i$ corresponds to a ghost field. The FP ghost term $S_{FP}$ represents non-physical elements (e.g., misinformation or bias) affecting opinion propagation among agents.

2. Mathematical Representation of FP Ghost Term:

$$S_{FP} = \int \bar{c}_i \mathcal{M}_{ij} c_j\, d^4x$$

Here, $\bar{c}_i$ and $c_j$ represent the ghost influences among agents, and $\mathcal{M}_{ij}$ is a matrix indicating the strength of these ghost influences.

## 9.2 Legendre Transformation and Renormalization Group Flow

1. Legendre Transformation: Perform a Legendre transformation on the action, including the FP ghost term, denoted as $S$, to obtain the generating function $Z(J)$. Derive the effective action $\Gamma(\phi)$ from the generating function.

2. Analysis of Renormalization Group Flow: Solve the renormalization group equations for the effective action $\Gamma(\phi)$ and analyze the flow. Confirm the occurrence of a Landau pole if the flow diverges to infinity and consider the possibility of renormalizability if it converges to a fixed point.

### 9.3 Analysis of Limit Cycles and Ergodicity

1. Detection of Limit Cycles: From the analysis of renormalization group equations, explore the potential formation of limit cycles in the system. If limit cycles exist, analyze ergodicity and the dynamic stability of opinions.

2. Local Potential Approximation and Loop Expansions: Simplify the dynamics of opinions using the local potential approximation. Perform loop expansions to examine the influence of higher-order interactions on opinion propagation.

### 9.4 Computational Process

Setting Initial Conditions: Assign opinions and the influence of the FP ghost term to agents.

Simulation and Analysis: Conduct simulations to calculate the dynamics of opinion propagation. Utilize Legendre transformation to derive the generating function and solve renormalization group equations. Analyze the presence of limit cycles and ergodicity to study opinion stability and variability.

It is important to note that interpreting and validating the application of quantum field theory equations in the context of social sciences pose challenges in this model.

## 10. Conclusion: Application of FP Ghost, Different Types of Cutoffs, York Decomposition, and UV Fixed Points in Social Science Models

FP ghosts, different types of cutoffs, York decomposition, and UV fixed points are concepts typically found in physics theories such as gravity theory and quantum chromodynamics. Applying these concepts directly to social science models is challenging and their theoretical validity remains unverified. However, let's attempt to apply these ideas at a hypothetical level to social science.

### 10.1 Introduction of FP Ghost Term

1. Definition of Ghost Term: The ghost term $S_{FP}$ models non-physical elements (e.g., misinformation or bias) in opinion propagation among agents. Formula: $S_{FP} = \int \bar{c}_i \mathcal{M}_{ij} c_j \, d^4x$ Here, $\bar{c}_i$ and $c_j$ represent the ghost influences among agents, and $\mathcal{M}_{ij}$ is the matrix indicating the strength of these influences.

2. Calculation of Ghost Term Influence: Add the influence of the ghost term to the opinion propagation among agents and calculate opinion updates. Update equation:

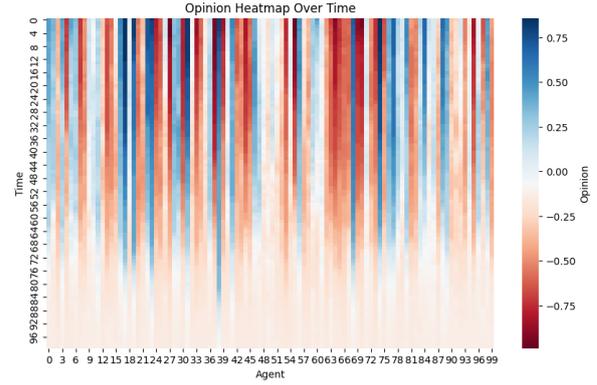

Fig. 14: Use $R_k^{Ia}$ Opinion Heatmap Over Time

$o_i^{(new)} = o_i + \alpha(o_j - o_i) + \gamma \bar{c}_i \mathcal{M}_{ij} c_j$ Here, $\alpha$ is the adaptation rate, and $\gamma$ is the strength of the ghost influence.

### 10.2 Application of Cutoff Functions

1. Definition of Cutoff Functions: Cutoff functions model constraints on opinion exchange. Formula: $R_k(o_i, o_j)$ This constrains interactions between opinions $o_i$ and $o_j$.

2. Calculations Based on Each Type of Cutoff: Type Ia Cutoff: Represents strong constraints. Use $R_k^{Ia}$. Type Ib Cutoff: Apply York decomposition for analysis of UV fixed points, use $R_k^{Ib}$. Type II Cutoff: Represents mild constraints. Use $R_k^{II}$. Type III Cutoff: Opinion exchange is active only under specific conditions. Use $R_k^{III}$.

### 10.3 Computational Process

1. Simulation Setup: Assign initial opinions $o_i$ to agents. Set the strength of ghost influence $\gamma$, adaptation rate $\alpha$, and cutoff function $R_k$.

2. Execution of Opinion Updates: At each time step, update opinions among agents. Use the update equation $o_i^{(new)} = o_i + \alpha(o_j - o_i) + \gamma \bar{c}_i \mathcal{M}_{ij} c_j$ and consider the influence of the cutoff function $R_k$.

3. Analysis of Results: Analyze opinion distribution, the presence of limit cycles, and ergodicity from simulation results.

  Type Ia cutoff: Represents strong constraints. Use $R_k^{Ia}$.

  Type Ib cutoff: Apply York decomposition. Use $R_k^{Ib}$ for the analysis of ultraviolet fixed points.

  Type II cutoff: Represents mild constraints. Use $R_k^{II}$.

  Type III cutoff: Active opinion exchange only under specific conditions. Use $R_k^{III}$.

### 10.4 Use $R_k^{Ia}$ of cutoff

(1) **Opinion Heatmap Over Time**:

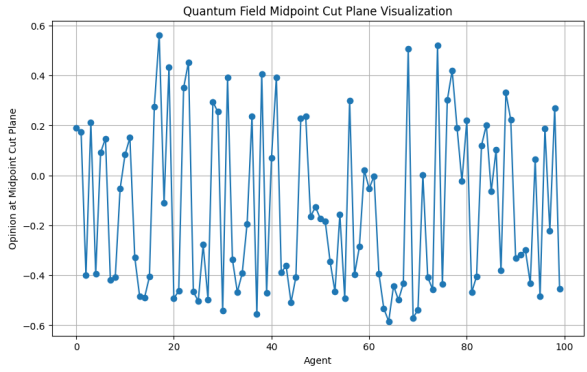

Fig. 15: Use $R_k^{Ia}$ Quantum Field Midpoint Cut Plane Visualization

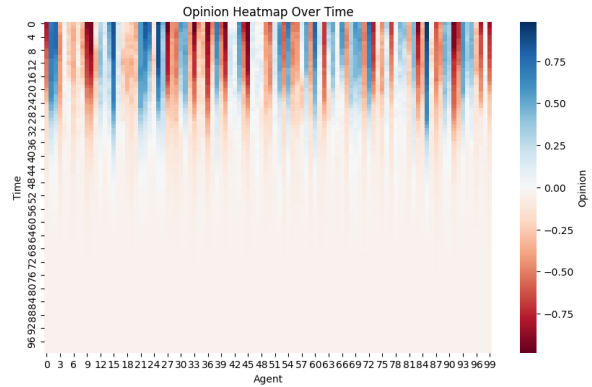

Fig. 16: Use $R_k^{Ib}$ Opinion Heatmap Over Time

This heatmap shows how opinions (ranging from -0.75 to 0.75) change over time across different agents (numbered 0 to 99).

The color gradient indicates the opinion value, with blue representing more negative opinions, red more positive ones, and white being neutral.

This visualization can be useful to observe the general trend in opinion changes, the emergence of consensus (if the colors become more homogenous), or the polarization (if we see the colors split into two distinct extreme groups).

(2) **Quantum Field Midpoint Cut Plane Visualization**:

This plot shows the opinion values for each agent at a specific point in time, likely the midpoint of the simulation.

The vertical lines represent the range of opinion values that each agent has held over time, while the dots indicate the opinion value at the midpoint.

A wide range of values for an agent suggests high volatility in their opinion, while a narrow range suggests a more stable opinion.

This plot can help identify which agents are most susceptible to opinion changes and which hold steadfast to their initial positions.

Given that you've mentioned the use of Ia-type cutoffs (representing strong constraints) using the $R_k^{Ia}$ function, we can infer that this particular analysis is looking at a scenario where agent interactions are significantly limited. In such cases, the ghost term's effect (representing non-physical elements like misinformation or bias) could either be exaggerated due to less frequent but more significant interactions or be dampened because the interactions are not frequent enough to propagate the anomalies widely. The precise interpretation would require more context about the parameters used in the simulation, the agents' initial opinion states, and the specific objectives of the analysis, such as whether it aims to reduce misinformation or study the effect of strongly enforced constraints on opinion dynamics.

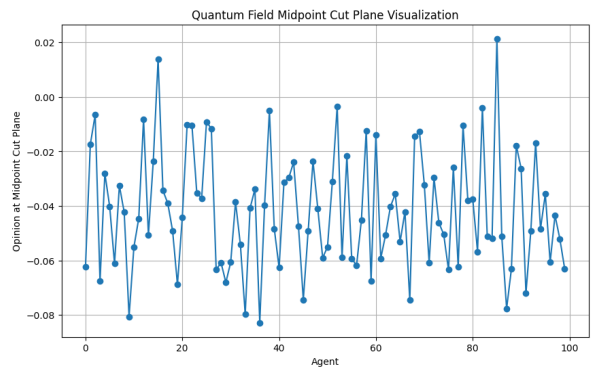

Fig. 17: Use $R_k^{Ib}$ Quantum Field Midpoint Cut Plane Visualization

## 10.5 Use $R_k^{Ib}$ of cutoff

### Heat Map Discussion

The heatmap shows the evolution of each agent's opinion with respect to time, with lighter colors (red tones) possibly representing more extreme or strong beliefs in opinions and darker colors (blue tones) representing milder opinions or neutral positions. If the color pattern changes over time, this suggests that the agent is influenced by external information (e.g., misinformation or information in a filter bubble) and that its opinion is changing. Such changes reflect the dynamics of information propagation in social interactions and provide clues to understanding how specific information sources and misinformation affect opinion formation.

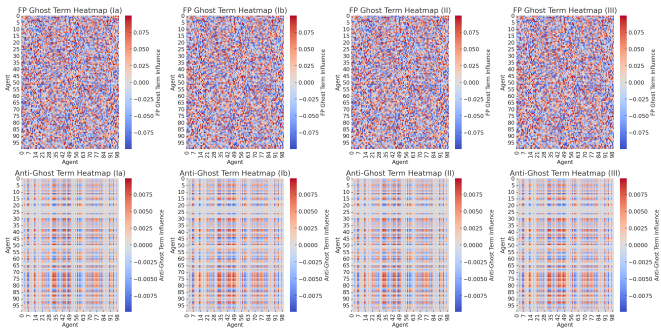

Fig. 18: FP-Ghost, Anti-Ghost Term Influence

## Consideration of Quantum Field Cutting Surfaces

Quantum field cut surfaces represent the state of an agent's opinion at a particular point in time. High peaks in value may indicate that an agent is highly influenced by misinformation or is in a self-reinforcing belief system (filter bubble). Low values, on the other hand, may indicate an agent who is less susceptible or accessing information from different sources. The distribution of peaks and valleys may indicate polarization of opinion or diversity of opinion within a society.

These graphs contribute to our understanding of the process of filter bubble formation and the impact of misinformation on social opinion. They may also help to elucidate the mechanisms by which misinformation and filter bubbles polarize opinions and how they act on the information ecosystem within society. In the social sciences, such analysis can help formulate strategies to reduce the spread of misinformation and create a more balanced information ecosystem.

## 10.6 FP Ghost Term, and Anti FP Ghost Term Heatmap discussion of FP ghost terms

The heatmap shows the influence of the ghost term between agents, represented by color intensity. Red can be interpreted as a strong influence and blue as a weak influence. The ghost terms model the non-physical elements (misinformation and prejudice) in the interactions between agents. The color patterns indicate how these elements are distributed across agents and may suggest that there is a particularly strong misinformation or prejudice influence among some agents.From this heatmap, we can analyze how misinformation and prejudice propagate in the network and contribute to the formation of filter bubbles.

## Anti-ghosting Term Heatmap Considerations

The anti-ghosting term heatmap may represent a countervailing factor to the influence of misinformation in interactions between agents. This could be interpreted as modeling the ability of agents to identify misinformation and suppress its impact. If the heatmap patterns are not uniform, it indicates that not all agents are able to identify misinformation to the same degree. This may be due to differences in education and information literacy.The distribution of the anti-ghosting terms allows us to analyze the strength and distribution of social immunity to misinformation. From these heat maps, we can understand the flow of information within a social network and how misinformation and bias affect opinion formation. They can also help identify which agents (or groups of people) are more susceptible to misinformation or more capable of suppressing it when developing social countermeasures. These insights have important implications when developing strategies to combat fake news and filter bubbles.

## 11. Conclusion:Modeling the Impact of Misinformation and Filter Bubbles

The case under consideration involves the transition where misconceptions of false information become eigenvalues believed as the ground state, leading to the eigenstates that only have the misconception of false information as eigenvalues. This can be formulated as integrals of the paths of the occurrence of false information or filter bubbles, resulting in a logic involving inference.

### 11.1 Introduction of Coordinate Operators in Heisenberg Picture

1. Application of Heisenberg Picture: Introduce coordinate operators $Q(t)$ in the Heisenberg picture, which correspond to social discussions and the flow of opinions. $Q(t)$ varies with time, reflecting the influence of misconceptions of false information.

### 11.2 Application of Operators Corresponding to Unlikely Information Propagation in Schrödinger Picture

1. Application of Schrödinger Picture: Define operators $\Psi$ in the Schrödinger picture, corresponding to the spread of fake news or unrealistic arguments. $\Psi$ represents states that share the same eigenvalues as false information.

2. Introduction of Time-Dependent Perturbations: Introduce perturbations that allow particles to stochastically evolve over time, representing the impact of false information. Time evolution is calculated using Green functions and taking the limit of infinite past.

### 11.3 Calculation Process

1. Impact on Ground State: Calculate the transition to states where only the ground state believes false news at the ground state $|0\rangle$. Integrals and inferences are calculated by taking the limit $t \to \infty$.

## 11.4 Analysis of Results

Analyze the mechanism of the impact of misconceptions of false information and the formation of filter bubbles from the calculated transition.

## 11.5 Mathematical Definitions

Time Evolution Operator $U(t,t') = T\exp\left(i\int_{t'}^{t} V(\tau)d\tau\right)$ Green Function $G(t,t') = i\langle 0|T[Q(t)Q(t')]|0\rangle$ Impact of Misconceptions of False Information $\Psi(t) = U^\dagger(t,\infty)Q(t)U(t,\infty)$

From here, the transition to states where misconceptions of false information are believed as eigenvalues in the ground state is formulated mathematically, and it is parameterized for scoring.

## 11.6 Mathematical Definitions

Impact of Misconceptions of False Information Operator: Define the operator $F$ representing the impact of misconceptions of false information. This operator represents the probability of agents believing fake news. $F$ acts on the opinion states of agents and increases the tendency to believe in fake news.

2. Impact on Ground State: In the ground state $|0\rangle$, a transition to states where misconceptions of false information are believed is induced by $F$. The probability amplitude of the transition is given by $\langle 0|F|0\rangle$.

3. Scoring Parameter: Introduce a scoring parameter $s$ to measure the magnitude of the impact of misconceptions of false information. $s$ is defined as the square of the absolute value of the transition probability amplitude $|\langle 0|F|0\rangle|^2$.

## 11.7 Scoring Calculation Process

Operator Setup: At the start of the simulation, set up the impact operator $F$ of false information.

Calculation of Transition Probability: Throughout the simulation, apply $F$ to the opinion states of each agent. Calculate the probability amplitude of the transition from the ground state to believing in false information.

Scoring: Calculate the score $s$ based on the probability of agents believing in false information. The score quantitatively represents the magnitude of the impact of false information.

## 11.8 Mathematical Definitions

Impact of Misconceptions of False Information Operator: $F$
Transition Probability Amplitude from Ground State: $\langle 0|F|0\rangle$
Scoring Parameter: $s = |\langle 0|F|0\rangle|^2$

Such a model offers a novel perspective on understanding the impact of false information in social science.

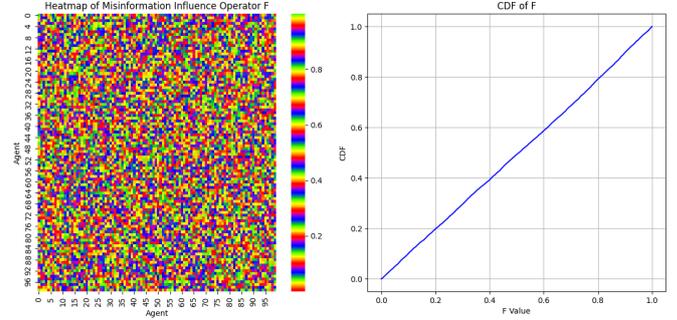

Fig. 19: CDF, Heatmap of Misinformation Influence Operator F

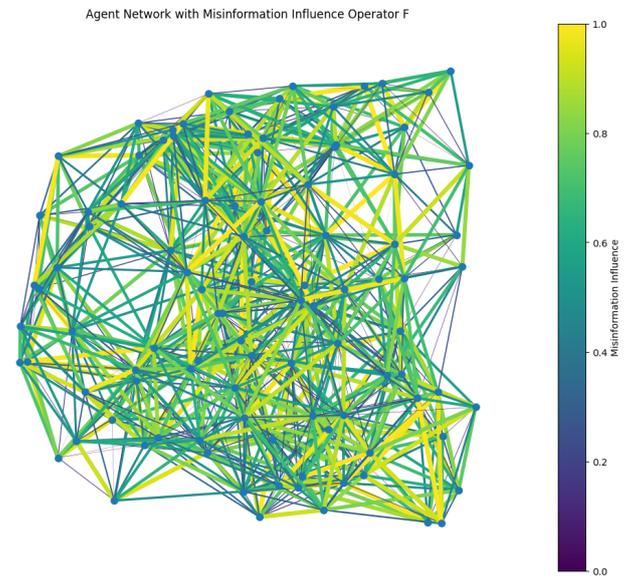

Fig. 20: Agent Network with Misinformation Influence Operator F

## 11.9 Heat Map Considerations

The heatmap shows the intensity of the influence operator $F$ of misinformation among agents, and the use of a variety of colors suggests that there is a large variation in the intensity of influence. The random pattern suggests that the misinformation is not randomly or evenly distributed within the network. This may indicate an environment in which misinformation is more likely to affect certain agents or groups. Such variation can help us understand how certain information influences agents' beliefs and behaviors in the formation of filter bubbles and the polarization of opinions.

## 11.10 CDF Considerations

The CDF shows how the values of the operator $F$ are distributed overall. A linear CDF shows that the values of $F$ are uniformly distributed, indicating that all agents are likely to be affected by misinformation to the same degree. The uniform distribution indicates the potential for widespread effects of misinformation within the social network and may represent a situation in which all agents are susceptible to misinformation.

## 11.11 Agent Network Visualization Considerations

The agent network visualization shows the strength of the influence of misinformation through the color and thickness of the edges. Edges that are brightly colored and thick mean that the influence of misinformation among those agents is high. Areas in the network with concentrations of strongly colored edges may indicate communities or groups where the influence of misinformation is particularly strong. This suggests an environment where extremes of opinion and filter bubbles are likely to form. These visualizations are useful for better understanding the propagation of misinformation and its effects in social networks. In particular, they provide insight into how misinformation affects individual agents and communities and how it contributes to opinion polarization and the formation of filter bubbles.

## 12. Conclusion: Information "Disconnection" and "Indefinite Metering Ghosts" in the Digital Environment

At the end of this discussion, the issue of information "disconnection" and "indefinite metering ghosts" in the digital environment Let us also consider this discussion in the context of hygiene issues in the digital environment.

Disconnection" of information and "indefinite metric ghosts" in the digital environment In the following discussion of patterns of disconnection, in the digital environment, disconnection refers to restrictions or obstacles in accessing information. There are several different patterns of disconnection. For example, access to the Internet may be cut off, access to certain information sources may be blocked, or information may be manipulated or altered. These disconnections would impede the free flow of information and potentially cause information gaps. Consideration would need to be given in the simulation regarding this social risk.

This is also an issue that is the subject of this paper. The problem of indefinite metric ghosting. Another issue is that the physical phenomenon of indefinite metric ghosting is not a very prominent information in the first place. In the digital environment, indefinite metric ghosting refers to reliability and security issues. When inaccurate, false, or malicious information exists and is mixed with reliable information, it becomes difficult for users to discern reliable information. This problem is an important issue regarding the quality and accuracy of information in the digital environment. Also, when considered in terms of the phase transition issue, phase transition in the digital environment refers to the phenomenon in which users have less access to different information and opinions on a given digital platform or social media, and are biased toward a particular source or point of view. This leaves users without access to diverse information, reinforces their own opinions and beliefs, and makes it difficult for them to access new information.

The same physical phenomenon of indefinite metric ghosting described above is manifested with respect to the existence of the Faddev-Popov ghost term (FP ghosting). In the digital environment, FP ghosts refer to "negative information" that gets mixed in with information. They exist as disinformation, hoaxes, or malicious information and can be confused with unreliable information. the presence of FP ghosts reduces the quality and reliability of information, making it difficult for users to find accurate information. However, this definition would allow for the detection of anomalous information on the quantum, as well as the detection factor of deteriorating quality information on the information, which could also be taken into account in the simulation for hypothesis and verification. To solve these problems, measures must be taken to improve the flow and quality of information in the digital environment. Digital users need to prevent information disconnections and develop the skills to identify reliable sources of information. Digital platforms and social media operators will also need to take measures to provide reliable information and eliminate inaccurate information and FP ghosting.

## Aknowlegement

The author is grateful for discussion with Prof. Serge Galam and Prof. Akira Ishii.